\documentclass[aps,prl,reprint,twocolumn,superscriptaddress,floatfix,nofootinbib,longbibliography]{revtex4-2}
\usepackage{graphicx,amsmath,amsfonts,amssymb,amsthm,xr}
\usepackage{epsfig}
\usepackage{tikz}
\usepackage{graphicx}
\usepackage{physics}
\usepackage{dcolumn}
\usepackage{bm}
\usepackage{ mathrsfs }
\usepackage{orcidlink}
\usepackage{algorithm}
\usepackage{algpseudocode}
\usepackage{float}

\usepackage{amsmath,amssymb}

\makeatletter
\renewcommand*\env@matrix[1][*\c@MaxMatrixCols c]{%
  \hskip -\arraycolsep
  \let\@ifnextchar\new@ifnextchar
  \array{#1}}
\makeatother

\usepackage{appendix} 

\begin{document}

\title{Mesons, Confinement, and Complexity: A Krylov Space Perspective on the Long-Range Ising Chain}

\author{Xuhao Jiang}
\affiliation{University of Regensburg, 93053 Regensburg, Germany}
\affiliation{Department of Physics and Arnold Sommerfeld Center for Theoretical Physics (ASC), Ludwig Maximilian University of Munich, 80333 Munich, Germany}
\affiliation{Technical University of Munich, TUM School of Natural Sciences, Physics Department, 85748 Garching, Germany}

\author{Jad C.~Halimeh${}^{\orcidlink{0000-0002-0659-7990}}$}
\affiliation{Department of Physics and Arnold Sommerfeld Center for Theoretical Physics (ASC), Ludwig Maximilian University of Munich, 80333 Munich, Germany}
\affiliation{Max Planck Institute of Quantum Optics, 85748 Garching, Germany}
\affiliation{Munich Center for Quantum Science and Technology (MCQST), 80799 Munich, Germany}
\affiliation{Department of Physics, College of Science, Kyung Hee University, Seoul 02447, Republic of Korea}

\author{N.~S.~Srivatsa${}^{\orcidlink{0000-0001-6433-450X}}$}
\email{srivatsans@iitj.ac.in}
\affiliation{Indian Institute of Technology Jodhpur, Jodhpur 342030, India}

\begin{abstract}

We investigate the interplay between confinement and Krylov complexity in the long-range transverse-field Ising chain, where power-law interactions $\sim 1/r^{\alpha}$ generate an effective confining potential between domain walls without an explicit symmetry-breaking field. Using the time-averaged Krylov state complexity after a quench as a diagnostic, we find that the complexity is strongly suppressed and only weakly system size-dependent at small $\alpha$ and small transverse field $h_x$, where confinement is strongest, and crosses over to rapid growth at larger $\alpha$, where it increases with system size. Within the confined regime, the oscillation frequencies of the complexity agree with the meson masses of an effective two-kink model, with the dominant frequency growing linearly with the lightest meson mass and the time-averaged complexity decreasing approximately as its inverse square. We further study operator Krylov complexity at finite temperature using the Wightman inner product. At low temperature, it exhibits bounded oscillations that directly encode the meson spectrum, whereas this structure is progressively washed out as the temperature increases. These results identify Krylov complexity as a spectroscopic probe of interaction-induced confinement and reproduce, in a microscopic lattice model, the hallmarks of Krylov complexity identified in confining field theories.

\end{abstract}

\maketitle

\textbf{\emph{Introduction.---}} Confinement, the binding of elementary excitations into composite particles that cannot propagate freely, is one of the most profound features of strongly coupled field theories~\cite{Weinberg_book,Gattringer2009QuantumChromodynamicsLattice,Zee2003QuantumFieldTheory,Rothe_book}. In quantum chromodynamics (QCD), quarks are permanently confined into hadrons by the linear confining potential of the chromoelectric flux tube stretching between them~\cite{Wilson1974}. An analogous mechanism emerges in quantum many-body systems, where domain walls play the role of quarks and are bound into mesons by the energy cost of the spin string connecting them~\cite{Calabrese1,Calabrese2,Calabrese3,McCoy,Liu2019,Lerose2019,Mussardo1,Mussardo2,Mussardo3,Mussardo4}, with signatures observed across a diverse range of experimental platforms~\cite{Wang2016,Morris2014,Grenier2015}.

\begin{figure}
    \includegraphics[width=\linewidth]{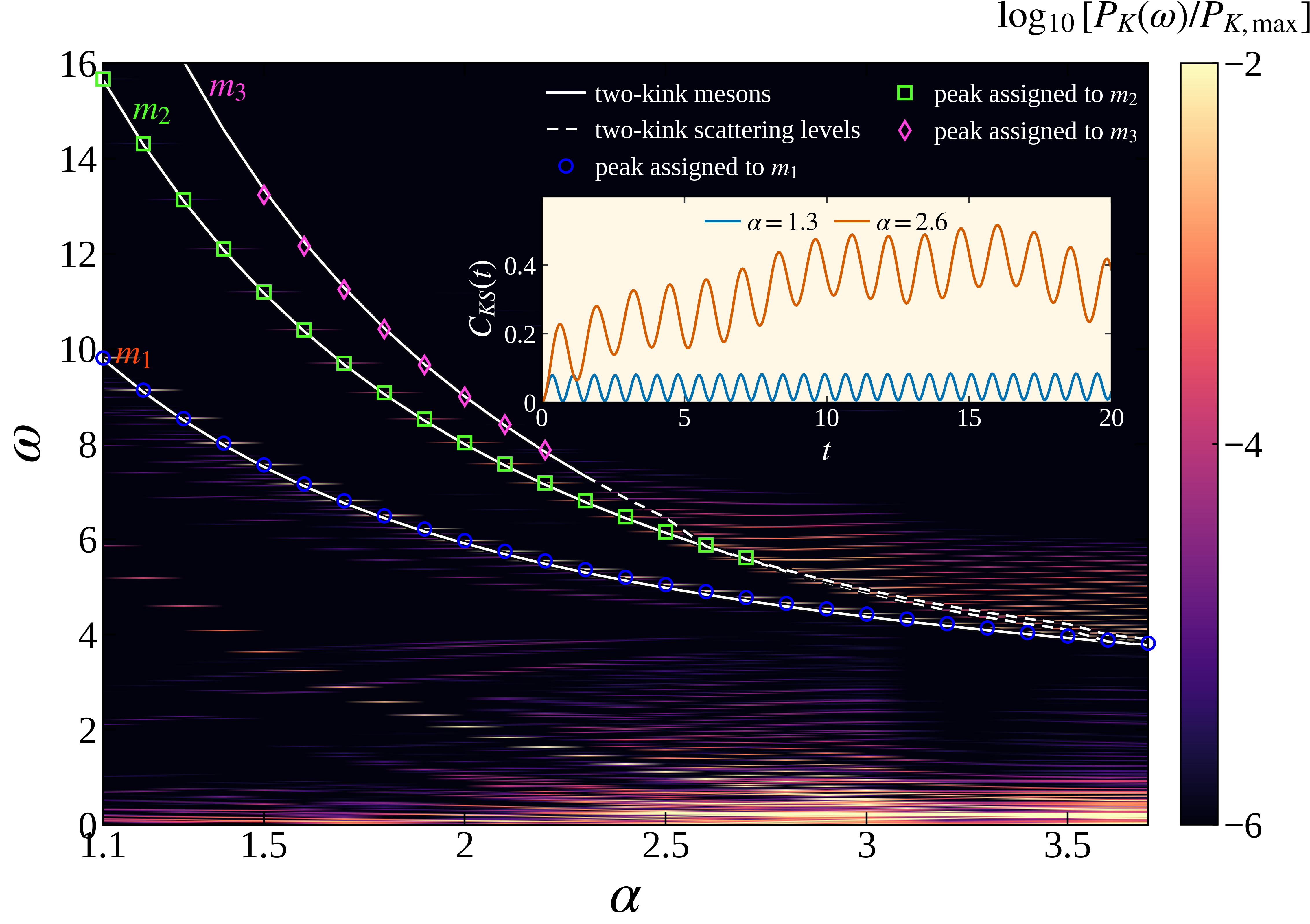}
    \caption{Fourier spectrum of the Krylov state complexity $C_{KS}(t)$ and meson masses in the long-range Ising model. The background shows $\log_{10}[P_K(\omega)/P_{K,\max}]$ for $L=16$ and $h_x=0.30$, with $P_K(\omega)$ defined in Eq.~(3). The discrete Fourier transform is performed over $0\leq t\leq 2000$ with a time step of $\Delta t=0.05$. White curves show the finite-size two-kink levels $m_{1,2,3}$~\cite{SM1}, solid where the level is a meson and dashed where it is a discretized scattering state, while colored markers indicate the spectral peaks assigned to the mesons; the weakest $m_3$ peaks lie below the color scale. The inset shows $C_{KS}(t)$ for two selected values of $\alpha$.}
    \label{f1}
\end{figure}

In the paradigmatic example, the Ising chain in a weak longitudinal field, the confining potential is generated by an explicit symmetry-breaking field~\cite{McCoy}. After a quench, confinement suppresses thermalization and entanglement spreading and induces oscillations that encode the meson spectrum~\cite{Kormos2017,Vovrosh2021}. Ising chains with couplings decaying as $1/r^{\alpha}$ display rich behavior beyond that of their nearest-neighbor counterparts~\cite{RevModPhys.95.035002,PhysRevLett.119.023001,PhysRevLett.121.090603,halimeh2017oct,halimeh2017dec,Halimeh2021sep,Halimeh2019jul}. Interestingly, they also provide a different source of confinement, in which the long-range tail alone binds domain walls into mesons without any symmetry-breaking field~\cite{Liu2019,Lerose2019}. For $\alpha\le2$ the energy of the spin string between two domain walls grows without bound with their separation, while for $\alpha>2$ it saturates~\cite{Liu2019}, so that $\alpha$ tunes confinement continuously. Such power-law interactions arise naturally in trapped-ion~\cite{islam2013,Tan2021} and Rydberg-atom~\cite{RevModPhys.82.2313,Bernien2017} quantum simulators.

\begin{figure*}
    \includegraphics[width=\linewidth]{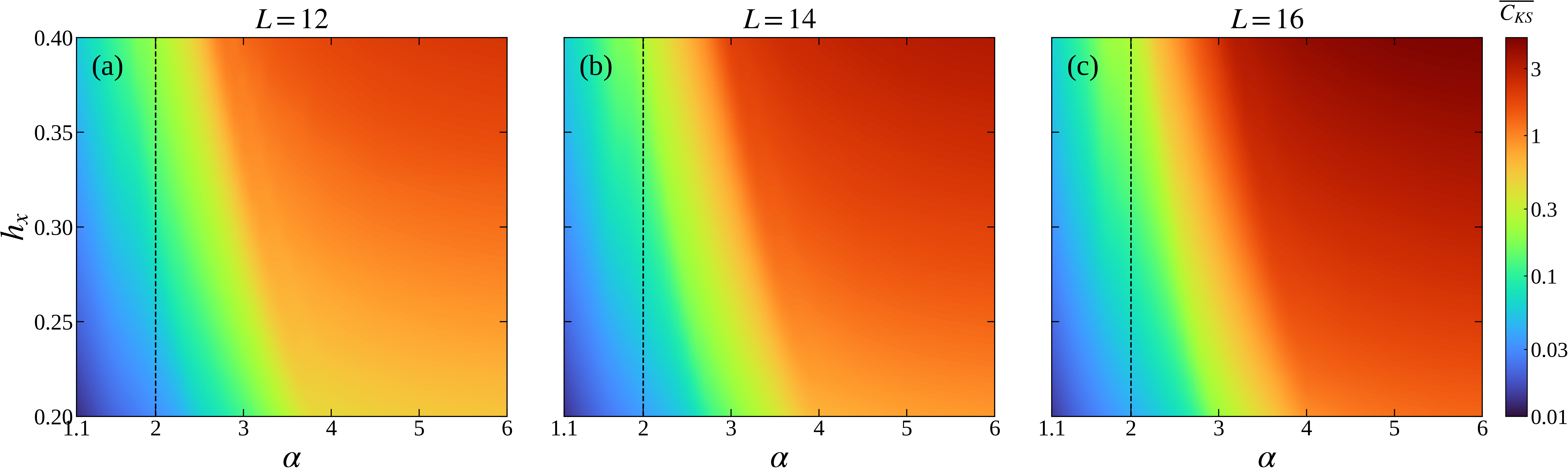}
    \caption{Time-averaged Krylov state complexity $\overline{C_{KS}}(\alpha,h_x)$ over the time window $t\in[0,200]$ following a quench from a fully polarized initial state under the long-range Hamiltonian $\hat{H}(\alpha,h_x)$ [Eq.~\eqref{eq:H}], shown on a logarithmic color scale for system sizes (a) $L=12$, (b) $L=14$, and (c) $L=16$. The dashed line marks $\alpha=2$, the threshold beyond which the long-range confining potential is known to saturate at large domain-wall separation~\cite{Liu2019}. For $\alpha\lesssim2$ the complexity is suppressed by up to two orders of magnitude and depends only weakly on system size $L$, whereas at large $\alpha$ it grows rapidly and increases with $L$.}
    \label{f2}
\end{figure*}

Krylov complexity has recently emerged as a probe of confinement \cite{xuhao2026,Fatemiabhari2026,Fatemiabhari2026b,Anegawa2024}. Starting from an initial state or operator, the Lanczos algorithm constructs an orthonormal Krylov basis through repeated action of the Hamiltonian or the Liouvillian, respectively. This maps the dynamics onto a particle hopping along a semi-infinite chain, and Krylov complexity, the mean position on this chain, measures how far the state or operator has spread~\cite{Parker2019,rabinovici,Vijay,Caputa2022}. It is known to serve as a complementary probe of quantum chaos in both closed and open quantum systems ~\cite{Parker2019,rabinovici,Hashimoto2023,Baggioli2025,Baggioli:2025knt,Chen2025,Alishahiha2025,Alishahiha2025b,gautam2023,NANDY20251,Cao_2021,Camargo2024,Rabinovici2022,Caputa:2025ucl,Rabinovici:2020ryf,Bhatta2022,Bhatta2025,Bhatta2023,Tanay2025,Sri2024,medina202,medina2025}. 

In large-$N$ gauge theories at finite temperature, operator Krylov complexity defined with the Wightman inner product has been proposed as an order parameter for the confinement--deconfinement transition. It oscillates in the confined phase, where the spectrum is discrete, and grows exponentially in the deconfined phase, where the spectrum is continuous~\cite{Anegawa2024}. Holographic studies likewise find that Krylov complexity oscillates in confining geometries, with amplitude and frequency set by the confinement scale~\cite{Fatemiabhari2026b,Fatemiabhari2026,jasper2026,Chatzis2026}. Likewise, in the short-range Ising chain with a longitudinal field, we showed that confinement induced by the longitudinal field suppresses the growth of Krylov complexity and induces oscillations whose frequencies encode the meson masses~\cite{xuhao2026}. Whether these signatures persist when confinement arises from long-range interactions in a lattice system, and how temperature affects them there, has not yet been explored. Nor is it understood to what extent such lattice results parallel the findings in confining field theories discussed above.

In this Letter, we address these questions in the long-range transverse-field Ising chain and observe features similar to several holographic constructions of Krylov complexity found in confining field theories \cite{Fatemiabhari2026,Fatemiabhari2026b,Anegawa2024}. After a quench, the Fourier spectrum of the state complexity resolves the two-kink meson masses over a wide range of $\alpha$, and its time average is strongly suppressed for $\alpha<2$ and crosses over to rapid growth at larger $\alpha$, where it increases with system size. As in holographic studies~\cite{Fatemiabhari2026b,Fatemiabhari2026}, stronger confinement raises the oscillation frequency and lowers the time-averaged complexity, following a similar scaling with the confinement scale. Introducing finite temperature through the Wightman inner product, we find that operator complexity at low temperature oscillates and directly probes the meson spectrum, while at high temperature this structure is washed out. This is reminiscent of the confined and deconfined phases of large-$N$ gauge theories~\cite{Anegawa2024}.

\textbf{\emph{Model.---}} 
\label{sec:model}
 We consider the one-dimensional long-range transverse-field Ising chain
\begin{equation}
\hat{H}(\alpha,h_x) = -\sum_{i<j}^{L} \frac{J}{r_{ij}^{\alpha}}\,\sigma_i^z \sigma_j^z \;-\; h_x \sum_{i=1}^{L} \sigma_i^x ,
\label{eq:H}
\end{equation}
where $\sigma_i^{\mu}$ are Pauli operators on site $i$, $L$ is the chain length, $J=1$ sets the energy scale, $h_x$ is the transverse field, and $\alpha$ controls the power-law decay of the Ising coupling. We impose periodic boundary conditions, with $r_{ij}$ the shortest distance between sites $i$ and $j$ on the ring, and restrict to $\alpha>1$ so that the thermodynamic limit is well defined \cite{Liu2019}. No longitudinal field is included, unlike the short-range case, where confinement of domain-wall kinks requires an explicit symmetry-breaking field such as a magnetic field in the $z$ direction \cite{Kormos2017}, here the long-range tail itself generates an effective attractive potential between kinks, producing confinement intrinsically as $\alpha$ is decreased \cite{Liu2019}. For $\alpha\to\infty$, Eq.~\eqref{eq:H} reduces to the nearest-neighbor Ising chain, exactly solvable via a Jordan-Wigner mapping, with a ferromagnetic-paramagnetic quantum critical point at $h_x=1$. Long-range interactions shift this transition to $h_x^{c}(\alpha)>1$, with $h_x^c(\alpha)\to1$ as $\alpha\to\infty$ \cite{Koffel2012,Fey2016}. Furthermore, a low-energy effective two-kink model shows that the long-range coupling alone confines pairs of domain walls~\cite{Liu2019} and for $\alpha \le2$ the confining potential grows without bound with the domain-wall separation, so all two-kink states are bound mesons. For $\alpha>2$ it saturates, so only the low-lying states are bound and a continuum of free domain-wall pairs lies above them.

In this work, we study how confinement is reflected in the growth of state complexity after a quench and of operator complexity in thermal equilibrium, using Krylov complexity, which we previously established as a complementary probe of confinement in Ref.~\cite{xuhao2026}.

\textbf{\textit{State complexity.}---} 
We first probe the imprint of long-range-induced confinement on the post-quench dynamics through Krylov state complexity~\cite{rabinovici}. Following a quench, the initial state $|\Psi_0\rangle$ generates the Krylov subspace $\mathcal{K}_n(\hat{H},|\Psi_0\rangle)=\mathrm{span}\{|\Psi_0\rangle,\hat{H}|\Psi_0\rangle,\ldots,\hat{H}^{n-1}|\Psi_0\rangle\}$ under the long-range Hamiltonian $\hat{H}(\alpha,h_x)$ of Eq.~\eqref{eq:H}. An orthonormal Krylov basis $\{|K_n\rangle\}$ is obtained via the Lanczos algorithm, which tridiagonalizes $\hat{H}$ as $\hat{H}|K_n\rangle=a_n|K_n\rangle+b_{n+1}|K_{n+1}\rangle+b_n|K_{n-1}\rangle$. Expanding the time-evolved state as $|\Psi(t)\rangle=\sum_n\psi_n(t)|K_n\rangle$, the Schr\"odinger equation reduces to a tight-binding problem on a semi-infinite chain, $i\partial_t\psi_n(t)=a_n\psi_n(t)+b_n\psi_{n-1}(t)+b_{n+1}\psi_{n+1}(t)$. The Krylov state complexity is then
\begin{equation}
C_{KS}(t)=\sum_n n\,|\psi_n(t)|^2 ,
\label{eq:CK_state}
\end{equation}
quantifying the spread of the wavefunction in Krylov space. We employ the constrained Lanczos algorithm ~\cite{Eckseler2025} throughout our paper which we detail in the Supplemental Material (SM)~\cite{SM1}. 

We first focus on the behavior of Krylov complexity in the confined regime which has been shown to be a sensitive indicator of confinement~\cite{xuhao2026}. We prepare the system in the state $|\Psi_0\rangle=|\uparrow\uparrow\cdots\uparrow\rangle$, fully polarized along $z$, and evolve it under the long-range Ising Hamiltonian $\hat{H}(\alpha,h_x)$ of Eq.~\eqref{eq:H} over a range of $\alpha$ fixing $h_x$ to be small. As shown in the inset of Fig.~\ref{f1}, the Krylov complexity $C_{KS}(t)$ exhibits characteristic oscillations that encode the meson masses, similar to those observed in Ref.~\cite{xuhao2026}. Following Ref.~\cite{xuhao2026}, we analyze the dynamics in the frequency domain through the power spectrum of the Krylov complexity,
\begin{equation}
P_{K}(\omega) = \left|\mathcal{F}\left[C_{KS}(t) - \overline{C_{KS}}\right]\right|^2,
\end{equation}
where $\mathcal{F}$ denotes the discrete Fourier transform over the simulated time window and $\overline{C_{KS}}$ is the time average of the Krylov complexity over the time window. As shown in Fig.~\ref{f1}, the power spectrum is peaked at the meson masses $m_{1,2,3}$ computed from the low-energy effective two-kink model of Ref.~\cite{Liu2019} (also detailed in the SM~\cite{SM1}), wherever these levels are mesons. The lightest meson dominates, while the peaks of the heavier mesons are much weaker. As $\alpha$ increases, the heavier branches successively lose their bound character and turn into discretized scattering states of the finite ring, whereas $m_1$ remains a meson over the whole range shown.

We next characterize quenches across a wider range of parameters $(h_x,\alpha)$ through the time average of the Krylov state complexity $\overline{C_{KS}}$ shown in Fig.~\ref{f2}. We average over a time window $t\in[0,200]$. In a finite chain, each state near $|{\uparrow\cdots\uparrow}\rangle$ and its spin-flipped partner near $|{\downarrow\cdots\downarrow}\rangle$ are split by an exponentially small tunneling gap $\delta$. On times $t\gtrsim1/\delta$, tunneling between them transfers weight to the far end of the Krylov chain and increases $C_{KS}(t)$, an effect absent in the thermodynamic limit. The window $t\in[0,200]$ is long compared with the meson oscillation periods and, over most of the parameter range, short compared with these tunneling times, so $\overline{C_{KS}}$ characterizes the symmetry-broken state. Only on the weakly confined side of the crossover does the lightest meson tunnel within the window, making $\overline{C_{KS}}$ a finite-window diagnostic~\cite{SM1}. For small $\alpha$ and $h_x$, we observe that $\overline{C_{KS}}$ is suppressed by up to two orders of magnitude and depends only weakly on system size. This suppression persists throughout $\alpha<2$, where the confining potential in the low-energy effective picture of Ref.~\cite{Liu2019} grows without bound. At larger $\alpha$, $\overline{C_{KS}}$ grows rapidly, signaling the weakening of confinement. The crossover, whose position is nearly independent of system size, moves to smaller $\alpha$ as $h_x$ increases, and beyond it $\overline{C_{KS}}$ increases with $L$.

\begin{figure}
    \includegraphics[width=\linewidth]{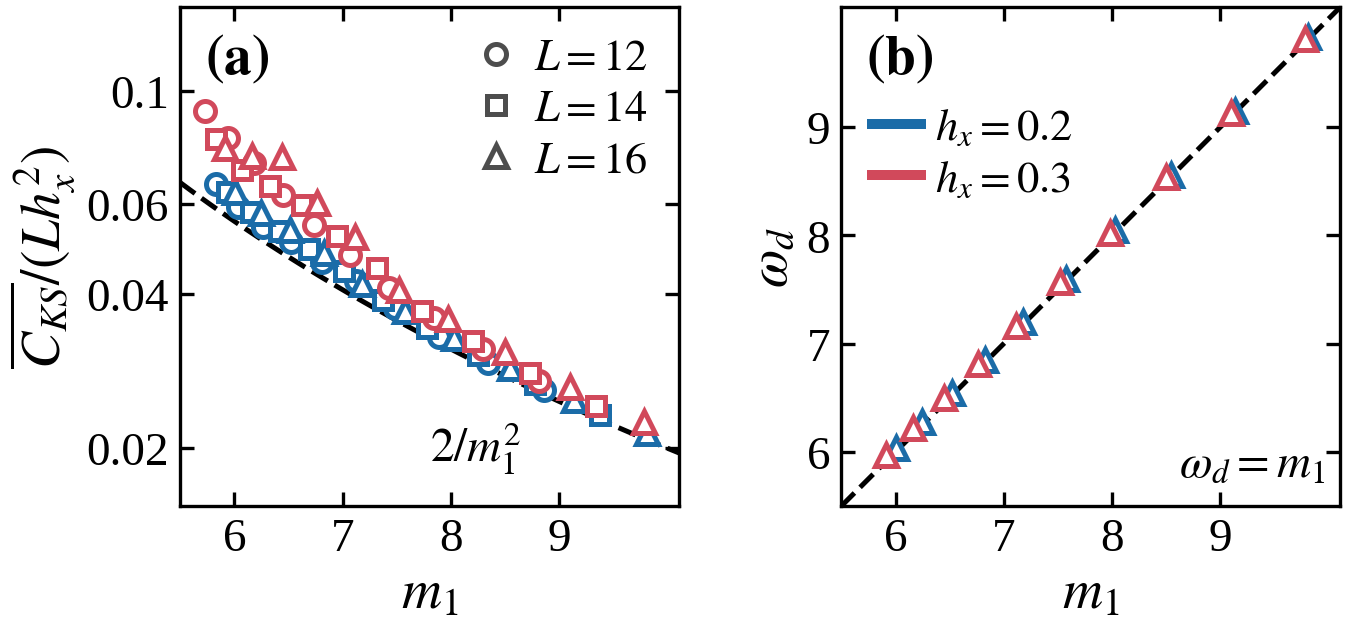}
    \caption{(a) Time-averaged Krylov state complexity $\overline{C_{KS}}$, scaled by $Lh_x^2$, versus the lightest meson mass $m_1$ for $L=12$, $14$, and $16$. The dashed line is the leading-order result $2/m_1^2$. (b) Dominant oscillation frequency $\omega_d$ of $C_{KS}(t)$ versus $m_1$ for $L=16$. The dashed line marks $\omega_d=m_1$. In both panels, $1.1\le\alpha\le2$, $h_x=0.2$ and $0.3$, and time averages are taken over the time window $t\in[0,200]$.}
    \label{f3}
\end{figure}

To quantify how confinement controls the dynamics in Krylov space, Fig.~\ref{f3} shows the time-averaged complexity $\overline{C_{KS}}$ and the dominant oscillation frequency $\omega_d$ for $1.1\le\alpha\le2$ as functions of the lightest meson mass $m_1$ computed independently from the two-kink approximation, whose dependence on $\alpha$ is set by the confining long-range interaction. Deep in the confined regime, the quench only weakly populates the first Krylov vector, coupled to the initial state by the first Lanczos coefficient $b_1=h_x\sqrt{L}$, giving to
leading order $\overline{C_{KS}}\simeq 2Lh_x^2/m_1^2$~\cite{SM1}. We therefore plot $\overline{C_{KS}}/(Lh_x^2)$, which removes the trivial dependence on $L$ and $h_x$ and isolates the dependence on the confinement scale. Indeed, $\omega_d$ coincides well with $m_1$ [Fig.~\ref{f3}(b)], and $\overline{C_{KS}}/(Lh_x^2)$ approaches $2/m_1^2$ at large $m_1$, with deviations at smaller $m_1$ reflecting the relative motion of the bound kinks [Fig.~\ref{f3}(a)]. This structure is analogous to holographic studies of confining field theories~\cite{Fatemiabhari2026}, where oscillations of Krylov complexity likewise arise as a consequence of confinement. There, the parameter $Q$ that sets the confinement scale controls these oscillations: the frequency grows as $\sqrt{Q}$, while the complexity
amplitude decreases as $1/Q$. Here $m_1$ plays the role of $\sqrt{Q}$, with $\omega_d\simeq m_1$ and $\overline{C_{KS}}\propto m_1^{-2}$.

\begin{figure*}
    \includegraphics[width=\linewidth]{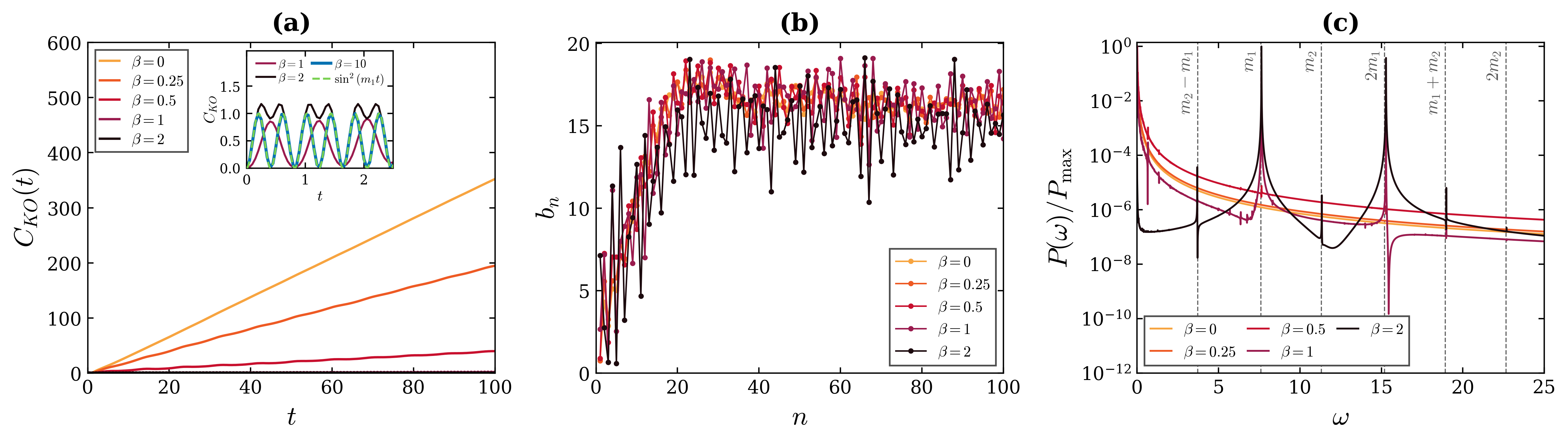}
    \caption{Operator Krylov dynamics of the domain wall-number operator $\hat{O}_{\mathrm{DW}}$ in the long-range transverse-field Ising chain with $L=13$, $h_x=0.3$, and $\alpha=1.4$, for a range of inverse temperatures $\beta$. The operator Krylov basis contains at most $K_{\max}=5000$ basis operators; numerical details are given in the Supplemental Material~\cite{SM1}. (a) Operator Krylov complexity $C_{KO}(t)$. Its approximately linear growth slows as $\beta$ increases. Inset: short-time view at large $\beta$, where $C_{KO}(t)$ oscillates and approaches the single-meson prediction $\sin^2(m_1 t)$. (b) Lanczos coefficients $b_n$ versus iteration $n$. Their staggering becomes more pronounced as $\beta$ increases, reflecting the gapped low-temperature spectral function. (c) Normalized power spectrum $P_O(\omega)/P_{O,\max}$, with $P_O(\omega)=|\mathcal{F}[C_{KO}(t)-\overline{C_{KO}}]|^2$ and $\mathcal{F}[\;]$ a discrete Fourier transform. Dashed lines mark the two-kink meson masses $m_{1,2}$ and their combinations $m_2-m_1$, $2m_1$, $m_1+m_2$, and $2m_2$. The discrete Fourier transform is performed over $0\leq t\leq 1000$ with a time step of $\Delta t=0.05$.}
    \label{f4}
\end{figure*}

\textbf{\textit{Operator complexity at finite temperature.}---}To probe confinement signatures independently of a fixed quench state, we study the Heisenberg growth of an initial operator $\hat{\mathcal{O}}_0$ under $\hat{H}(\alpha,h_x)$, regularized at finite temperature via the Wightman inner product, $(\hat{A}|\hat{B})_\beta = \frac{1}{Z}\,\mathrm{Tr}\!\left[e^{-\beta \hat{H}/2}\hat{A}^\dagger e^{-\beta \hat{H}/2}\hat{B}\right]$,
with $Z=\mathrm{Tr}[e^{-\beta\hat H}]$, which reduces to the infinite-temperature (Frobenius) inner product as $\beta\to0$. Our use of the Wightman inner product follows Ref.~\cite{Anegawa2024}, where temperature drives a confinement--deconfinement transition in large-$N$ gauge theories. Here, by contrast, confinement is set by $\alpha$, and temperature instead controls which transitions dominate the operator dynamics. Applying the Lanczos algorithm to the Liouvillian $\hat{\mathcal{L}}\hat{O}=[\hat{H},\hat{O}]$ with respect to $(\cdot|\cdot)_\beta$ generates an orthonormal operator basis $\{\hat{O}_n\}$ and Lanczos coefficients $b_n(\beta)$, with $\hat{\mathcal{O}}(t)=\sum_n \varphi_n(t)\hat{O}_n$ evolving as a semi-infinite hopping problem analogous to the state case. The operator Krylov complexity is $C_{KO}(t)=\sum_n n\,|\varphi_n(t)|^2$ computed similar to the state complexity.

We first show that operator Krylov complexity serves as a direct spectroscopic probe of the meson spectrum, with the resolution sharpening as temperature is lowered. The central object is the normalized Wightman autocorrelation function $G_\beta(t) = (\hat{O}_0|e^{i\hat{\mathcal{L}}t}\hat{O}_0)_\beta$. Its Lehmann representation shows that the peak positions are determined by many-body energy differences, with spectral weights that depend on both temperature and the choice of operator. At low temperature, transitions from the ground state $|0\rangle$, with energy $E_0$, to the meson bound states $|M_j\rangle$ with energies $E_{M_j}$ dominate the spectral weight, and $G_\beta(t)\simeq w_0+\sum_j p_j\cos(m_j t)$ with $p_j\propto|F_j|^2e^{-\beta m_j/2}$. Here $m_j=E_{M_j}-E_0$ are the meson excitation energies, $F_j=\langle M_j|O|0\rangle$ are their form factors, and $w_0$ is the zero-frequency spectral weight. Since the ground state is a doublet split by an exponentially small tunneling gap, $|0\rangle$ denotes either of its two members, which contribute equally at temperatures above this splitting. The thermal factor $e^{-\beta m_j/2}$, inherited from the Wightman inner product, exponentially suppresses heavier mesons with increasing $\beta$, so that as $\beta \to \infty$ only the lightest meson with a nonzero form factor survives, while the zero-frequency weight $w_0$, carried by thermally excited states, is suppressed as well. In this single-mode limit the operator Krylov complexity takes the exact form $C_{KO}(t) = \sin^2(m_1t)$, oscillating at twice the lowest meson mass $m_1$ with a long-time average of $1/2$ independent of $m_1$. However, we note that this form is generic to any spectral function dominated by a single isolated peak; here, its frequency directly yields the lightest meson mass, in agreement with the two-kink bound-state energy~\cite{SM1}. Details of the calculation are provided in the SM~\cite{SM1}. More generally, its Fourier spectrum contains peaks at $2m_j$ and $|m_i\pm m_j|$, and, when zero-frequency weight is present, at $m_j$ itself~\cite{SM1}, providing a direct imprint of the meson states accessible to the operator. At higher temperature, transitions between thermally excited states acquire spectral weight. Their frequencies, set by energy differences between many-meson states, fill in around the meson lines and wash out the isolated peaks. 

To test this, we consider the zero-momentum domain-wall-number operator, $\hat{O}_{\mathrm{DW}}= \frac{1}{\sqrt{L}}\sum_{j=1}^{L}\frac{1-\sigma_j^{z}\sigma_{j+1}^{z}}{2}$, which measures the total kink content of the spin configuration and, up to an additive constant, is equivalent to the nearest-neighbor Ising bond-energy operator. We restrict our finite-temperature analysis to the confined regime, $\alpha<2$ and small $h_x$, where finite-size effects are weakest. We first study the behavior of the Lanczos coefficients $b_n(\beta)$ as shown in Fig.~\ref{f4}(b). At high temperature they rise over the first few tens of Lanczos iterations $n$ before saturating at a plateau set by the finite system size. As $\beta$ increases, the coefficients develop a pronounced even--odd staggering. This staggering is the Krylov-space signature of a discrete, gapped spectral function where at low temperature the spectral weight of $\hat{O}_{\mathrm{DW}}$ concentrates near $\omega=0$ and $\pm m_j$, and in the single-meson limit the Lanczos algorithm terminates after $b_1=m$, $b_2=0$, the extreme case of staggering [see Eq.~\eqref{eq:op_single_meson} in the SM~\cite{SM1}]. Alternating strong and weak hopping amplitudes dimerize the Krylov chain and hinder the operator from spreading along it. Such staggering accompanies any gapped spectral function; here the gap is set by the lightest meson.

The consequences are directly visible in the operator complexity shown in Fig.~\ref{f4}(a). At high temperature, $C_{KO}(t)$ grows approximately linearly in time, as expected for saturated Lanczos coefficients. This growth slows markedly as $\beta$ increases, and the complexity develops oscillations that approach the single-meson form $C_{KO}(t)=\sin^2(m_1 t)$ (inset). The power spectrum in the frequency domain shown in Fig.~\ref{f4}(c) makes the meson content explicit. At low temperature its peaks coincide with the two-kink masses and their combinations, whereas at high temperature transitions between excited states populate many additional frequencies and the discrete structure is progressively washed out.

These observations are analogous to large-$N$ gauge theories~\cite{Anegawa2024}, where operator complexity oscillates in the confined phase and grows exponentially in the de-confined phase. Here, lowering the temperature likewise turns the approximately linear growth of $C_{KO}(t)$ over the simulated time window into bounded oscillations set by the meson masses. In our finite chains this change is a crossover controlled by which transitions carry the Wightman weight, and its relation to thermodynamic phase transitions remains open.

\textbf{\emph{Conclusions.---}} We have shown that Krylov complexity provides a sensitive probe of confinement induced purely by long-range interactions in
the transverse-field Ising chain. After a quench, the oscillations of the state complexity resolve the meson masses over a range of $\alpha$. Its time average is suppressed by up to two orders of magnitude for $\alpha<2$ and crosses over to rapid growth at larger $\alpha$, where it increases with system size. Within the confined regime, stronger confinement raises the oscillation frequency, set by the lightest meson mass, and lowers the time-averaged complexity, mirroring findings from recent studies in holographic confining field theories~\cite{Fatemiabhari2026,Fatemiabhari2026b}. At finite temperature, operator complexity defined with the Wightman inner product oscillates at low temperature in the confined regime, with spectral peaks at the meson masses and their combinations and pronounced even--odd staggering of the Lanczos coefficients. Both features are progressively washed out as the temperature increases. This behavior is reminiscent of that found in confining large-$N$ gauge theories~\cite{Anegawa2024}. 

These signatures are encoded in spectral functions. The state complexity is fixed by the return amplitude $\langle\Psi_0|e^{-i\hat Ht}|\Psi_0\rangle$ and the operator complexity by the autocorrelation function $G_\beta(t)$, whose moments determine the Lanczos coefficients through the recursion method~\cite{Parker2019}. Both are dynamical correlation functions that are accessible, at least indirectly, in quantum simulators~\cite{Baez2020}. Although reconstructing a long sequence of Lanczos coefficients from finite, noisy data is ill-conditioned, the leading signatures identified here, the oscillation at the lightest meson mass and the suppression of the state complexity, are fixed by the first Lanczos coefficients and should be the most robust to extract.

Moving forward, several questions remain open. A natural next step is to understand how confinement is reflected in the growth of Krylov complexity in lattice gauge theories, where confinement arises from the interplay of gauge and matter fields~\cite{Weinberg_book,Gattringer2009QuantumChromodynamicsLattice,Zee2003QuantumFieldTheory,Rothe_book,Yaffe1980,Mildenberger2025,Zhang2023,Halimeh2022tuning}. These theories also host string breaking, in which a gauge-flux string breaks through the creation of dynamical matter~\cite{Surace2020,mwy1-v9hk}. It would be interesting to explore how Krylov complexity behaves during such a string-breaking event, and how these regimes are encoded in the Lanczos coefficients and in the complexity growth of quantum states and operators. Furthermore, Krylov complexity at finite temperatures may shed light on how thermal fluctuations affect confinement in these theories~\cite{Gavai:1994bt,Kebri2024}.

\bigskip
\footnotesize
\begin{acknowledgments}
\textbf{\textit{Acknowledgments.---}}The authors acknowledge Cheuk Yiu Wong for fruitful discussions and for valuable insights from related work that contributed to this project. The authors acknowledge funding by the Max Planck Society, the Deutsche Forschungsgemeinschaft (DFG, German Research Foundation) under Germany’s Excellence Strategy – EXC-2111 – 390814868, and the European Research Council (ERC) under the European Union’s Horizon Europe research and innovation program (Grant Agreement No.~101165667)—ERC Starting Grant QuSiGauge.
\end{acknowledgments}
%


\clearpage
\widetext
\begin{center}
\textbf{\large Supplemental material for `Mesons, Confinement, and Complexity: A Krylov Space Perspective on the Long-Range Ising Chain'}
\textbf{\large }
\end{center}
\makeatletter
\renewcommand{\c@secnumdepth}{0}
\makeatother
\setcounter{equation}{0}
\setcounter{figure}{0}
\setcounter{table}{0}
\setcounter{page}{1}
\makeatletter
\renewcommand{\theequation}{S\arabic{equation}}
\renewcommand{\thefigure}{S\arabic{figure}}
\renewcommand{\thesection}{S\arabic{section}}

\section{Operator complexity and the meson spectrum}

We now examine how meson excitations appear in operator Krylov complexity. Throughout this work, site indices are understood modulo \(L\), and we use the thermally connected components
\(\delta O=O-\langle O\rangle_{\beta}\mathbb{I}\), normalized with respect to the Wightman inner product, as the initial Krylov operators. For a Hermitian operator, define $\mathcal N_\beta=(\delta O|\delta O)_\beta>0$ and $\hat O_0=\delta O/\sqrt{\mathcal N_\beta}$. The normalized Wightman autocorrelation function is
\begin{equation}
G_\beta(t)=(\hat O_0|e^{i\hat{\mathcal L}t}\hat O_0)_\beta
=\int d\omega\,\rho_\beta(\omega)e^{i\omega t}.
\label{eq:op_G_definition}
\end{equation}
Here $\rho_\beta(\omega)$ is the normalized spectral function. Inserting energy eigenstates $\hat H|a\rangle=E_a|a\rangle$ into the trace gives its Lehmann representation. Using $\hat{\mathcal L}(|a\rangle\langle b|)=(E_a-E_b)|a\rangle\langle b|$ and $\delta O_{ab}=\langle a|\delta O|b\rangle$, we obtain

\begin{equation}
\rho_\beta(\omega)=\frac{1}{Z\mathcal{N}_\beta}\sum_{a,b}e^{-\beta(E_a+E_b)/2}\,|\delta O_{ab}|^2\,\delta\bigl(\omega-(E_a-E_b)\bigr).
\label{eq:op_Lehmann}
\end{equation}

Thus $\rho_\beta\geq0$ and $\int\rho_\beta\,d\omega=G_\beta(0)=1$. Exchanging $a,b$ and using Hermiticity gives $\rho_\beta(-\omega)=\rho_\beta(\omega)$. The peak positions are set by many-body energy differences, while their spectral weights depend on temperature and the initial operator.

At low temperature, suppose that the finite-frequency spectral weight is dominated by transitions from the zero-momentum ground-state doublet to meson states $|M_j\rangle$. Its two members, related by spin-flip parity and split by an exponentially small tunneling gap, contribute equally at temperatures above this gap; we denote either of them by $|0\rangle$. Their masses are $m_j=E_{M_j}-E_0>0$, and their form factors are $F_j=\langle M_j|O|0\rangle$. Combining the $(M_j,0)$ and $(0,M_j)$ contributions in Eq.~\eqref{eq:op_Lehmann} gives

\begin{equation}
G_\beta(t)\simeq w_0+\sum_j p_j\cos(m_j t),\qquad p_j=\frac{2e^{-\beta E_0}}{Z\mathcal{N}_\beta}\,|F_j|^2e^{-\beta m_j/2}.
\label{eq:op_meson_G}
\end{equation}

Here $w_0$ is the zero-frequency spectral weight, obtained by summing the terms with $E_a=E_b$, and $w_0+\sum_jp_j\simeq1$ within this approximation. The factor two accounts for the contributions at positive and negative frequencies. The exponent contains $m_j/2$ because the Wightman inner product places a thermal factor on each side of the operator. Thus $p_j/p_i=|F_j/F_i|^2e^{-\beta(m_j-m_i)/2}$ for these ground-state transitions with $F_i\neq0$. Lowering the temperature increases the relative weight of the lowest-mass meson with a nonzero form factor. However, the zero-frequency contribution and transitions involving nearly degenerate low-energy states may still be important.

The domain-wall-number operator, $\hat{O}_{\mathrm{DW}}= \frac{1}{\sqrt{L}}\sum_{j=1}^{L}\frac{1-\sigma_j^{z}\sigma_{j+1}^{z}}{2}$ discussed in the main text conserves momentum and global spin-flip parity $P=\prod_j\sigma_j^x$. The corresponding selection rules require $|M_j\rangle$ to have the same momentum and parity as the ground-state member it connects to for $F_j$ to be nonzero, and a meson contributes no weight if its form factor vanishes. The meson masses can be estimated independently using the two-kink approximation at weak transverse field~\cite{Liu2019}, whereas the spectral weights also require the form factors. At higher temperature, transitions between excited states in Eq.~\eqref{eq:op_Lehmann} can produce additional peaks at meson energy differences.

To derive the spectral decomposition of $C_{KO}(t)$, consider the Liouvillian matrix $\mathsf L_{nn'}=(\hat O_n|\hat{\mathcal L}\hat O_{n'})_\beta$ in the full operator Krylov subspace of the finite system. Let $\lambda_s$ and $u_{ns}$ denote its eigenvalues and orthonormal eigenvector components. The initial condition $\varphi_n(0)=\delta_{n0}$ gives $\varphi_n(t)=\sum_su_{ns}u_{0s}^{*}e^{i\lambda_st}$. Substituting into the definition of $C_{KO}$ yields

\begin{equation}
C_{KO}(t)=\sum_{r,s}A_{rs}\,e^{i(\lambda_s-\lambda_r)t},\qquad A_{rs}=u_{0r}u_{0s}^{*}\sum_n n\,u_{nr}^{*}u_{ns}.
\label{eq:op_complexity_spectrum}
\end{equation}

If $\rho_\beta$ consists of peaks at $0,\pm m_j$, the Fourier spectrum of $C_{KO}$ can contain peaks at $m_j$, $2m_j$, $|m_i-m_j|$, and $m_i+m_j$, provided their coefficients are nonzero. These frequencies are differences of Liouvillian eigenvalues and need not coincide with meson masses.

Consider a single meson mode with no zero-frequency contribution, described by $\rho(\omega)=[\delta(\omega-m)+\delta(\omega+m)]/2$ with $m>0$. Then $\hat{\mathcal L}^{2}\hat O_0=m^2\hat O_0$. Since the spectral function is even, $\hat O_1=\hat{\mathcal L}\hat O_0/m$ is normalized and orthogonal to $\hat O_0$. Moreover, $\hat{\mathcal L}\hat O_1=m\hat O_0$, so the Lanczos algorithm terminates after two basis operators ($b_1=m$, $b_2=0$). Time evolution in this two-dimensional Krylov subspace gives
\begin{equation}
\begin{aligned}
\hat O_0(t)&=\cos(mt)\hat O_0+i\sin(mt)\hat O_1,\\
G(t)&=\cos(mt),\qquad C_{KO}(t)=\sin^2(mt).
\end{aligned}
\label{eq:op_single_meson}
\end{equation}
Thus the autocorrelation function oscillates at $m$, while the complexity oscillates at $2m$ and has a long-time average of $1/2$, independent of $m$. With a zero-frequency contribution, $\rho(\omega)=w_0\delta(\omega)+\frac{1-w_0}{2}[\delta(\omega-m)+\delta(\omega+m)]$, the Lanczos algorithm terminates after three basis operators, with $b_1=m\sqrt{1-w_0}$ and $b_2=m\sqrt{w_0}$, giving $C_{KO}(t)=(1-w_0)\sin^2(mt)+8w_0(1-w_0)\sin^4(mt/2)$. The zero-frequency weight therefore adds a peak at $m$ to the complexity spectrum, which dominates over the peak at $2m$ when $w_0>0.1$; Eq.~\eqref{eq:op_single_meson} is recovered as $w_0\to0$. The dominant peak at $m_1$ for $\beta=2$ in Fig.~\ref{f4}(c) of the main text reflects such a residual zero-frequency weight. Additional meson modes can give sum and difference frequencies, as described by Eq.~\eqref{eq:op_complexity_spectrum}.

In the low-temperature regime of Eq.~\eqref{eq:op_meson_G}, the meson contribution is therefore characterized by frequencies fixed by the excitation energies and spectral weights determined by the form factors and temperature. Lowering the temperature enhances the relative weight of the lowest-mass meson with a nonzero form factor, while changing the initial operator can redistribute the weight among the same meson modes. 

\section{Finite-size meson masses in the two-kink approximation}
\label{sec:finite_size_mesons}

The finite-size two-kink spectrum provides a common mass reference for Krylov state and operator complexity. A meson is a bound pair of domain walls: the Ising interaction supplies the confining potential, while the transverse field moves the walls. Following Liu \textit{et al.}~\cite{Liu2019}, we describe this motion through the position and length of a reversed spin segment. We formulate the two-kink Hamiltonian on the same periodic ring as the full spin Hamiltonian in Eq.~\eqref{eq:H},
\[
\hat H=\hat H_z-h_x\sum_{j=1}^{L}\sigma_j^x,
\qquad
\hat H_z=-J\sum_{a<b}\frac{\sigma_a^z\sigma_b^z}{r_{ab}^{\alpha}},
\qquad
r_{ab}=\min(|a-b|,L-|a-b|),
\]
where $J>0$, $\alpha>1$, and site indices are understood modulo $L$. The projected spectrum estimates the meson energies; the complexity dynamics in the main text is generated by the full spin Hamiltonian.

\textit{Two-kink basis and interaction energy.---}
Let $|F_\uparrow\rangle=|\uparrow\cdots\uparrow\rangle$ be the fully polarized state. Reversing $n$ consecutive spins gives $|j,n\rangle=\prod_{\ell=0}^{n-1}\sigma_{j+\ell}^x|F_\uparrow\rangle$, where $j$ labels the first reversed spin and $n=1,\ldots,L-1$. Each configuration has a domain wall at either end of the segment. With $\mathcal P_2$ the projector onto these states, the two-kink Hamiltonian is $\hat H_2=\mathcal P_2(\hat H-E_F)\mathcal P_2$, where $E_F=\langle F_\uparrow|\hat H_z|F_\uparrow\rangle$ sets the energy reference.

A bond contributes to the segment energy when exactly one of its spins is reversed, changing its energy from $-J/r_{ab}^{\alpha}$ to $+J/r_{ab}^{\alpha}$. Summing over bonds that connect the reversed set $S_{j,n}=\{j,\ldots,j+n-1\}$ to its complement gives
\begin{equation}
V_L(n)=\langle j,n|\hat H_z|j,n\rangle-E_F
=2J\sum_{a\in S_{j,n}}\sum_{b\notin S_{j,n}}
\frac{1}{r_{ab}^{\alpha}}.
\label{eq:meson_finite_potential}
\end{equation}
A segment and its complement have the same crossing bonds, so $V_L(n)=V_L(L-n)$. This expression includes the finite-ring geometry at every separation. For fixed $n$ and $L\to\infty$, it approaches the infinite-chain potential of Ref.~\cite{Liu2019},
\[
V_\infty(n)=4J\left[n\zeta(\alpha)
-\sum_{r=1}^{n-1}\frac{n-r}{r^{\alpha}}\right],
\]
where $\zeta$ is the Riemann zeta function.

\textit{Relative motion of the domain walls.---}
A spin flip at either boundary increases or decreases the segment length by one. Projecting these processes onto the two-kink basis yields

\begin{equation}
\hat H_2|j,n\rangle=V_L(n)|j,n\rangle-h_x\bigl(|j-1,n+1\rangle+|j,n+1\rangle+|j+1,n-1\rangle+|j,n-1\rangle\bigr).
\label{eq:twokink_action}
\end{equation}

with terms of length $0$ or $L$ excluded. Translation symmetry separates the overall motion of the pair from the relative coordinate $n$. In the momentum basis
\begin{equation}
|k,n\rangle=\frac{1}{\sqrt L}\sum_{j=1}^{L}
e^{-ik(j+n/2)}|j,n\rangle,
\qquad k=\frac{2\pi q}{L},\quad q=0,\ldots,L-1,
\label{eq:meson_momentum_basis}
\end{equation}
the left and right boundary moves contribute phases $e^{-ik/2}$ and $e^{ik/2}$. Their sum gives the hopping amplitude $-2h_x\cos(k/2)$. The resulting $(L-1)\times(L-1)$ matrix, $[\mathsf H_2(k)]_{nn'}=\langle k,n|\hat H_2|k,n'\rangle$, is
\begin{equation}
[\mathsf H_2(k)]_{nn'}=V_L(n)\delta_{nn'}
-2h_x\cos(k/2)\bigl(\delta_{n',n+1}+\delta_{n',n-1}\bigr).
\label{eq:meson_finite_matrix}
\end{equation}
The kink-separation wavefunction therefore obeys

\begin{equation}
V_L(n)\chi_{\nu,k}(n)-2h_x\cos(k/2)\bigl[\chi_{\nu,k}(n+1)+\chi_{\nu,k}(n-1)\bigr]=\varepsilon_\nu(k)\chi_{\nu,k}(n),\qquad \chi_{\nu,k}(0)=\chi_{\nu,k}(L)=0.
\label{eq:meson_finite_eigenproblem}
\end{equation}

The endpoint conditions exclude the two fully polarized configurations, which contain no domain walls. A bound pair has a wavefunction concentrated at small separation $\min(n,L-n)$, and $\varepsilon_\nu(k)$ gives its dispersion within the two-kink approximation.

\textit{Symmetry-resolved mass estimates.---}
At $k=0$, global spin flip $P=\prod_j\sigma_j^x$ acts as $P|0,n\rangle=|0,L-n\rangle$, where the first ket entry denotes momentum. The symmetry $V_L(n)=V_L(L-n)$ separates $\mathsf H_2(0)$ into parity sectors $p=\pm1$, with orthonormal bases
\begin{equation}
|n,p\rangle=\frac{|0,n\rangle+p|0,L-n\rangle}{\sqrt2},
\qquad 1\leq n<L/2.
\label{eq:meson_even_basis}
\end{equation}
For even $L$, the state $|0,L/2\rangle$ completes the $p=+$ basis. If $R_p$ contains the basis vectors as columns, diagonalizing $R_p^\dagger\mathsf H_2(0)R_p$ gives the ordered eigenvalues $\varepsilon_{\nu,p}(0)$. We denote the corresponding mass estimates by $m_{\nu,p}^{(2)}(L)=\varepsilon_{\nu,p}(0)$. Their reference energy $E_F$ is the classical ferromagnetic vacuum energy, giving the weak-transverse-field two-kink estimate of the excitation gaps~\cite{Liu2019}. Coupling to the zero- and higher-kink sectors generates the vacuum and excitation-energy corrections beyond this order.

\textit{Reference masses for the state complexity.}---The fully polarized initial state has zero momentum and contains both spin-flip parities, $|F_\uparrow\rangle=(|F_+\rangle+|F_-\rangle)/\sqrt2$ with $|F_\pm\rangle=(|F_\uparrow\rangle\pm|F_\downarrow\rangle)/\sqrt2$, so both parity sectors enter its dynamics. The transverse field couples it directly to the shortest reversed segment, $(\hat H-E_F)|F_\uparrow\rangle=-h_x\sqrt{L}\,|0,1\rangle$. In a weak quench, interference between vacuum-like and meson components then produces oscillations of $C_{KS}(t)$ at the meson excitation energies, with weights set by the initial-state overlaps, and interference between meson components adds difference frequencies. The two-kink levels $m^{(2)}_{\nu,p}(L)$ provide the leading estimate of these energies; for bound mesons the two parities coincide up to exponentially small tunneling corrections. The reference branches $m_{1,2,3}$ in Fig.~\ref{f1} are the lowest $k=0$ two-kink levels, retained only if they are separated by more than the frequency resolution $\Delta\omega=2\pi/t_{\max}$ of the spectrum.

\textit{Two-state benchmark for the state complexity.---}
For the polarized initial state $|K_0\rangle=|F_\uparrow\rangle$, as noted above, the first Lanczos step is exact: $|K_1\rangle=|0,1\rangle$, with $b_1=h_x\sqrt{L}$ and $a_1-a_0=V_L(1)$, the energy cost of one reversed spin. If the dynamics is dominated by a single tightly bound meson, the Krylov chain reduces, up to a constant energy shift, to the two-state Hamiltonian in the basis $\{|K_0\rangle,|K_1\rangle\}$ with coupling $g$ and detuning $\Delta$,
\begin{equation}
\hat H_2=\begin{pmatrix}0 & g\\ g & \Delta\end{pmatrix},\qquad
C(t)=\frac{4g^2}{\Delta^2+4g^2}\sin^2\!\left(\frac{\sqrt{\Delta^2+4g^2}}{2}\,t\right),
\end{equation}
with time average $\overline{C}=2g^2/(\Delta^2+4g^2)$. Setting $g=h_x\sqrt{L}$ and $\Delta\simeq m_1\gg g$ gives $\omega_d\simeq m_1$ and $\overline{C_{KS}}\simeq2Lh_x^2/m_1^2$, the trends compared in Fig.~\ref{f3}. This is a benchmark for the weakly coupled regime, not an exact infinite-time result for the full system.

\textit{Reference masses for the operator complexity.}---By the selection rules of Sec.~S1, transitions from the even (odd) member of the zero-momentum ground-state doublet access the $k=0$, $p=+$ ($p=-$) mesons, whose masses coincide for bound states up to exponentially small tunneling corrections. The reference masses in Fig.~\ref{f4} are therefore the ascending eigenvalues $m^{(2)}_{\nu,+}(L)$, evaluated at the system size and couplings of that calculation, and the same spectrum supplies the sum and difference frequencies of $C_{KO}(t)$ [Eq.~\eqref{eq:op_complexity_spectrum}]. Thermally populated states contribute further transitions within their own momentum and parity sectors.

\section{Constrained Lanczos algorithm for states and operators}
\label{sec:constrained_lanczos}

The initial state or operator selects the part of the spectrum that participates in its dynamics. We use this spectral support to construct the corresponding Krylov subspace before applying the Lanczos algorithm. This constrained representation follows the approach of Eckseler \textit{et al.}~\cite{Eckseler2025}, who showed that finite-precision Lanczos iterations can acquire components outside the exact Krylov subspace, for both state and operator calculations. Expressing the problem directly in a basis of that subspace keeps the Lanczos vectors within it by construction. Below we give the state and operator versions for the full spin Hamiltonian.

\textit{Krylov state complexity.---}
For a normalized state $|\Psi_0\rangle$, the Krylov subspace is $\mathcal K_S=\operatorname{span}\{\hat H^\ell|\Psi_0\rangle\}_{\ell\geq0}$. To identify it, write $\hat H=\sum_\mu E_\mu P_\mu$, where the $E_\mu$ are distinct energies and $P_\mu$ projects onto the entire eigenspace at energy $E_\mu$. The component of the initial state in that eigenspace has norm $\sqrt{p_\mu}$ and normalized direction $|u_\mu\rangle$:
\begin{equation}
\begin{aligned}
p_\mu&=\langle\Psi_0|P_\mu|\Psi_0\rangle,
& |u_\mu\rangle&=\frac{P_\mu|\Psi_0\rangle}{\sqrt{p_\mu}}
\quad (p_\mu>0),\\
\hat H^\ell|\Psi_0\rangle&=\sum_\mu E_\mu^\ell\sqrt{p_\mu}\,|u_\mu\rangle.
\end{aligned}
\label{eq:constrained_state_projection}
\end{equation}
All sums below run over the nonzero projections. Different $|u_\mu\rangle$ are orthonormal because they belong to distinct energy eigenspaces. Within a degenerate eigenspace, repeated action of $\hat H$ simply multiplies the initial projection by $E_\mu$; it therefore generates only this one direction.

Equation~\eqref{eq:constrained_state_projection} shows that every Krylov vector is a linear combination of the $|u_\mu\rangle$. Conversely, each of these directions can be isolated by a polynomial in $\hat H$. In particular, the interpolation polynomial
\[
f_\mu(x)=\prod_{\nu\neq\mu}\frac{x-E_\nu}{E_\mu-E_\nu}
\quad\text{satisfies}\quad
f_\mu(\hat H)|\Psi_0\rangle=\sqrt{p_\mu}\,|u_\mu\rangle.
\]
Thus, $\{|u_\mu\rangle\}$ forms an orthonormal basis of $\mathcal K_S$, whose dimension $K_S$ is the number of distinct energies with nonzero initial-state weight. We perform the Lanczos iteration in this space without introducing a separate cutoff on the Krylov dimension.

With $U=(|u_1\rangle,\ldots,|u_{K_S}\rangle)$, the Hamiltonian and initial state in this basis are
\begin{equation}
\begin{aligned}
D_S&=U^\dagger\hat H U=\operatorname{diag}(E_\mu),
& (q_0^S)_\mu&=\sqrt{p_\mu},\\
e^{-i\hat Ht}|\Psi_0\rangle&=Ue^{-iD_St}q_0^S.
\end{aligned}
\label{eq:constrained_state_representation}
\end{equation}
The pair $(D_S,q_0^S)$ therefore contains the complete dynamics selected by $|\Psi_0\rangle$. The Lanczos algorithm transforms this spectral basis into the Krylov basis $\{|K_n\rangle\}$ used to define $C_{KS}(t)$.

\textit{Operator Krylov complexity.---}
The operator construction has the same structure, with the Hamiltonian replaced by the Liouvillian $\hat{\mathcal L}A=[\hat H,A]$. In the energy eigenbasis, $\hat{\mathcal L}(|a\rangle\langle b|)=(E_a-E_b)|a\rangle\langle b|$, so the relevant eigenvalues are energy differences rather than energies.

At inverse temperature $\beta$, we use the Wightman inner product~\cite{Parker2019},
\[
(A|B)_\beta=\frac{1}{Z}\operatorname{Tr}
\left(e^{-\beta \hat H/2}A^\dagger e^{-\beta \hat H/2}B\right),
\qquad Z=\operatorname{Tr}e^{-\beta \hat H}.
\]
The initial operator is the normalized connected component $\hat O_0=\delta O/\sqrt{\mathcal N_\beta}$, where $\delta O=O-\langle O\rangle_\beta\mathbb I$, $\langle O\rangle_\beta=Z^{-1}\operatorname{Tr}(e^{-\beta \hat H}O)$, and $\mathcal N_\beta=(\delta O|\delta O)_\beta>0$. To incorporate the thermal metric into the operator components, define
\begin{equation}
\begin{aligned}
\mathcal S_\beta(A)&=Z^{-1/2}e^{-\beta \hat H/4}Ae^{-\beta \hat H/4},\\
\operatorname{Tr}\!\left[\mathcal S_\beta(A)^\dagger\mathcal S_\beta(B)\right]
&=(A|B)_\beta,
\qquad
\mathcal S_\beta([\hat H,A])=[\hat H,\mathcal S_\beta(A)].
\end{aligned}
\label{eq:constrained_thermal_map}
\end{equation}
The first identity follows from cyclicity of the trace; the second follows because $e^{-\beta \hat H/4}$ commutes with $\hat H$. The map therefore preserves both inner products and Liouvillian evolution. We can consequently use the ordinary Hilbert--Schmidt (Frobenius) inner product for $\hat X_0=\mathcal S_\beta(\hat O_0)$, whose matrix elements are
\[
(\hat X_0)_{ab}=\frac{e^{-\beta(E_a+E_b)/4}\delta O_{ab}}
{\sqrt{Z\mathcal N_\beta}},
\qquad \sum_{a,b}|(\hat X_0)_{ab}|^2=1.
\]

For each distinct energy difference $\omega_\mu$ present in $\hat X_0$, combine the matrix elements with that frequency into a single normalized direction:
\begin{equation}
\begin{aligned}
w_\mu&=\sum_{E_a-E_b=\omega_\mu}|(\hat X_0)_{ab}|^2,\\
\hat B_\mu&=\frac{1}{\sqrt{w_\mu}}
\sum_{E_a-E_b=\omega_\mu}(\hat X_0)_{ab}|a\rangle\langle b|,
\qquad w_\mu>0.
\end{aligned}
\label{eq:constrained_frequency_projection}
\end{equation}
These operators obey $\operatorname{Tr}(\hat B_\mu^\dagger \hat B_\nu)=\delta_{\mu\nu}$ and $\hat{\mathcal L}\hat B_\mu=\omega_\mu \hat B_\mu$, while $\hat X_0=\sum_\mu\sqrt{w_\mu}\hat B_\mu$. Just as in the state construction, all components sharing one eigenvalue evolve with the same phase and contribute one direction. Applying the preceding polynomial argument gives a Krylov dimension $K_O$ equal to the number of distinct gaps with nonzero weight. The constrained representation and its autocorrelation are therefore
\begin{equation}
\begin{aligned}
D_O &= \operatorname{diag}(\omega_\mu), \qquad (q_0^O)_\mu = \sqrt{w_\mu},\\
G_\beta(t) &= (q_0^O)^\dagger e^{iD_O t}\, q_0^O = \sum_\mu w_\mu\, e^{i\omega_\mu t}.
\end{aligned}
\label{eq:constrained_representation}
\end{equation}
A nonzero weight at $\omega_\mu=0$ is included in this sum and produces a time-independent contribution. The connected subtraction makes $\delta O$ orthogonal to the identity, while other matrix elements within equal-energy eigenspaces can retain zero-frequency weight.

\textit{Lanczos algorithm and Krylov dynamics.---}
Both constructions now reduce to a real diagonal matrix $D$ and a normalized initial vector $q_0$. Starting from $q_{-1}=0$ and $c_0=0$, the Lanczos algorithm generates orthonormal vectors through the three-term recurrence
\begin{equation}
\begin{aligned}
d_n&=q_n^\dagger Dq_n,
& r_n&=Dq_n-d_nq_n-c_nq_{n-1},\\
c_{n+1}&=\|r_n\|,
& q_{n+1}&=r_n/c_{n+1}.
\end{aligned}
\label{eq:constrained_recursion}
\end{equation}
At each step, the components along the current and preceding Krylov vectors are subtracted, and the remaining vector defines the next direction. The coefficients are $(d_n,c_n)=(a_n,b_n)$ for both states and operators. They form the diagonal and nearest-neighbor entries of the tridiagonal Krylov matrices $\mathsf H_{nn'}=\langle K_n|\hat H|K_{n'}\rangle$ and $\mathsf L_{nn'}=(\hat O_n|\hat{\mathcal L}\hat O_{n'})_\beta$, respectively. The latter is the Liouvillian matrix used in Sec.~S1. In exact arithmetic, the algorithm terminates when all supported directions have been generated: $b_{K_S}=0$ or $b_{K_O}=0$.

For a Hermitian seed, $\hat X_0$ is Hermitian, so exchanging $a$ and $b$ in Eq.~\eqref{eq:constrained_frequency_projection} gives equal weights at $\omega$ and $-\omega$. The Lanczos vectors can be written as $(q_n)_\mu=\sqrt{w_\mu}\,\pi_n(\omega_\mu)$, where $\pi_n$ is the degree-$n$ orthonormal polynomial for these weights. Symmetry of the weights gives $\pi_n(-\omega)=(-1)^n\pi_n(\omega)$, and hence
\[
a_n=\sum_\mu w_\mu\omega_\mu|\pi_n(\omega_\mu)|^2=0.
\]
The operator chain therefore has vanishing diagonal coefficients. The state chain retains $\alpha_n$, which encodes its generally asymmetric energy distribution.

The initial state or operator occupies the first Krylov site, represented by $\boldsymbol e_0=(1,0,\ldots)^{\mathsf T}$. Evolution on the two chains and the corresponding mean Krylov positions are
\begin{equation}
\begin{aligned}
\boldsymbol\psi(t)&=e^{-i\mathsf Ht}\boldsymbol e_0,
& C_{KS}(t)&=\sum_{n=0}^{K_S-1}n|\psi_n(t)|^2,\\
\boldsymbol\varphi(t)&=e^{i\mathsf Lt}\boldsymbol e_0,
& C_{KO}(t)&=\sum_{n=0}^{K_O-1}n|\varphi_n(t)|^2.
\end{aligned}
\label{eq:constrained_dynamics}
\end{equation}
The opposite signs in the propagators follow from Schr\"odinger state evolution and Heisenberg operator evolution with $\hat{\mathcal L}=[\hat H,\cdot]$. In either case, the complexity measures the mean distance traveled from the initial site along its Krylov chain.

In the numerical implementation, nearby energies or gaps are grouped as follows. For energies, the grouping tolerance is $\epsilon_E\max(1,\max_\mu|E_\mu|)$, with $\epsilon_E=\max(10^{-12},10^{-10}\|\hat H\|_F)$ for $L=12,14$ and $\epsilon_E=10^{-14}$ for $L=16$; here $\|\hat H\|_F$ is the Frobenius norm of the full Hamiltonian. The state calculation retains the projected initial-state direction of each energy group with $\sqrt{p_\mu}>10^{-12}$. The operator calculation retains components with $|(\hat X_0)_{ab}|>10^{-13}$ and groups adjacent sorted gaps whose separation does not exceed $10^{-10}\max(1,\max_{ab}|E_a-E_b|)$, where the maximum is taken over the retained components. It sums the weights within each gap group and uses their weight-averaged frequency $\omega_\mu$, without an additional group-weight cutoff. The retained initial vector is then normalized. These operations give a finite-resolution implementation of the spectral construction. For Krylov state complexity, we use the Full Orthogonalization Lanczos (FOL) algorithm with two orthogonalization passes at each step within the retained space; the iteration terminates when the residual norm falls below $10^{-12}$ or the retained spectral support is exhausted. For the operator calculations, the Lanczos algorithm retains at most $K_{\max}=5000$ basis operators, with earlier termination when the residual norm falls below $10^{-12}\max(1,\max_\mu|\omega_\mu|)$ or the retained spectral support is exhausted. Time evolution is evaluated by diagonalizing the resulting finite tridiagonal Hamiltonian or Liouvillian matrix, respectively.

\section{Choice of the averaging window}
Here we describe the choice of the time window $t\in[0,T]$, with $T=200$, used to compute the time-averaged Krylov state complexity shown in Fig.~\ref{f2}. The fully polarized initial state $|\Psi_0\rangle$ has equal weight on both members of each spin-flip parity doublet: a state near $|{\uparrow\cdots\uparrow}\rangle$ and its partner near $|{\downarrow\cdots\downarrow}\rangle$, split by a tunneling gap $\delta$ that is exponentially small in $L$. On times $t\gtrsim1/\delta$, tunneling between the two polarized sectors therefore produces a slow oscillation of $C_{KS}(t)$, a finite-size effect that is absent in the thermodynamic limit. A strict $T\to\infty$ average at fixed $L$ would include this tunneling and satisfy $\overline{C_{KS}}\geq1-\sum_E p_E^2\geq1/2$, where $p_E$ is the weight of $|\Psi_0\rangle$ in the eigenspace of energy $E$. We instead average over a window $t\in[0,200]$, which is long compared with the meson oscillation periods but short compared with the tunneling times, so that $\overline{C_{KS}}$ captures the plateau reached after the quench.

\begin{figure}
\includegraphics[width=10cm]{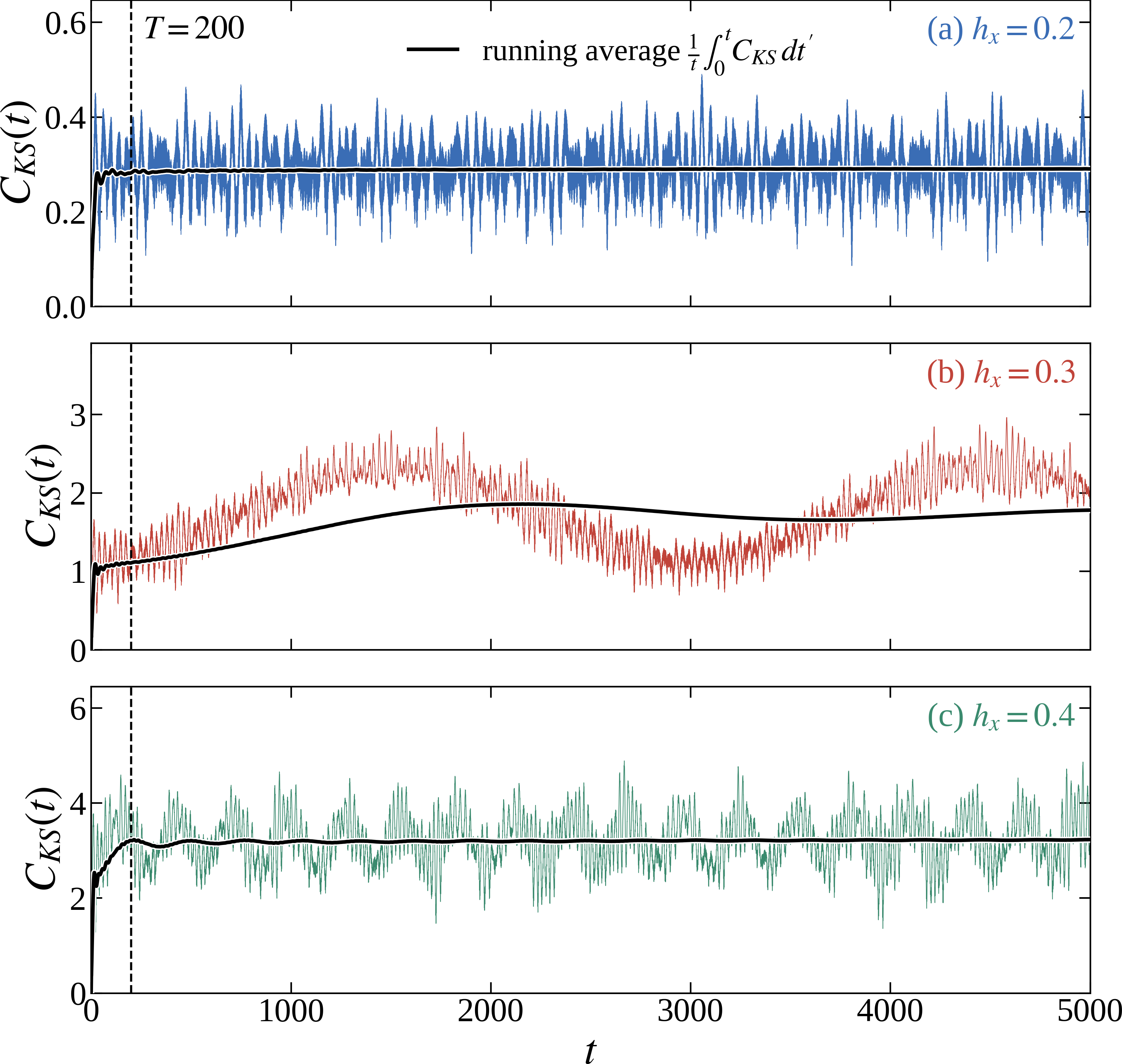}
\caption{Krylov state complexity $C_{KS}(t)$ for $L=16$ and $\alpha=3.2$ at (a) $h_x=0.2$, (b) $h_x=0.3$, and (c) $h_x=0.4$. The dashed line marks the averaging window $T=200$. The thick black lines denote the running average $t^{-1}\int_0^t C_{KS}\,dt'$. The slow oscillations in (b) and (c), with periods of about $3\times10^3$ and $3\times10^2$, arise from tunneling of the weakly bound lightest meson between its configurations on the two polarized backgrounds.}\label{fs1}
\end{figure}

Figure~\ref{fs1} illustrates this for $L=16$ and $\alpha=3.2$. For $h_x=0.2$, $C_{KS}(t)$ oscillates rapidly about a mean that stays constant over the entire simulated time, so its average does not depend on $T$. For $h_x=0.3$, the mean remains approximately constant up to $t\approx200$ but then rises and returns with a period of about $3\times10^3$, the signature of coherent tunneling of the weakly bound lightest meson between the two polarized backgrounds; the window $T=200$ excludes this finite-size effect. For $h_x=0.4$, the tunneling period is about $3\times10^2$, so the window already spans most of a period and cannot separate the tunneling from the fast oscillations; the average then hardly depends on $T$. These parameters lie on the weakly confined side of the crossover. Deep in the confined regime, by contrast, the tunneling periods exceed the simulated times by many orders of magnitude, while far beyond the crossover the tunneling is fast and averages out within the window and in both cases $\overline{C_{KS}}$ is independent of $T$, and the choice of window matters only in a band on the weakly confined side of the crossover. The running averages $t^{-1}\int_0^t C_{KS}\,dt'$ in Fig.~\ref{fs1} make this explicit: in (a) and (c) they are constant beyond $t\approx200$, whereas in (b) tunneling raises the average for longer windows.

\end{document}